\documentclass{article}
\usepackage{authblk}
\usepackage{graphicx}

\title{Exploring Spatial Coherence in Inter-annual Changes and Annual Extremes of Rainfall over India}
\author[1]{Adway Mitra}
\author[2]{Ashwin K. Seshadri}
\affil[1]{International Center for Theoretical Sciences, Bangalore, India}
\affil[2]{Indian Institute of Science, Bangalore. India}

\begin{document}
\maketitle	

\begin{abstract}
Forecasts of monsoon rainfall for India are made at national scale. But there is spatial coherence and heterogeneity that is relevant to forecasting. This paper considers year-to-year rainfall change and annual extremes at sub-national scales. We use Data Mining techniques to gridded rain-gauge data for 1901-2011 to characterize coherence and heterogeneity and identify spatially homogeneous clusters. We study the direction of change in rainfall between years (Phase), and extreme annual rainfall at both grid level and national level. Grid-level Phase is found to be spatially coherent, and significantly correlated with all-India mean rainfall (AIMR) phase. Grid-level extreme-rainfall years are not strongly associated with corresponding extremes in AIMR, although in extreme AIMR years local extremes of the same type occur with higher spatial coherence. Years of extremes in AIMR entail widespread phase of the corresponding sign. Furthermore, local extremes and phase are found to frequently co-occur in spatially contiguous clusters.
\end{abstract}	
	
\section{Introduction}
Forecasting of seasonal rainfall, especially the summer monsoon, is important to the Indian economy (\cite{Gadgil2010,Mall2006,Selvaraju2007}). Seasonal forecasts of rainfall are made at the national-scale (\cite{Rajeevan2007,Sahai2003}) because monsoons are large scale phenomena and there is an association between all-India summer monsoon rainfall and aggregate impacts (\cite{Kumar2011}). However rainfall is a spatially heterogeneous phenomenon, and the country may be divided into distinct homogeneous rainfall zones, based on mean rainfall (\cite{Srinivas2013,Gadgil1980}). There are also many regional differences in inter- and intra-annual variability (\cite{Azad2010,Gadgil2003,Goswami2005,Rajeevan2010}), rainfall trends and the occurrence of extreme events (\cite{Ghosh2009,Ghosh2012}). Apart from the South-West monsoon winds affecting major parts of the country and causing rainfall during the months June-September (JJAS), other factors play a role in monsoon rainfall [\cite{Gadgil2003}]. These include the retreating monsoon rainfall on the eastern coast particularly during October and November [\cite{Rajeevan2010}], and the Western Disturbances affecting North-western parts of the country during summer months [\cite{Gadgil2003}]. Furthermore, orography plays an important role [\cite{Gadgil2003}].

This paper studies spatial heterogeneity in interannual differences and extremes of rainfall, for both individual grid-points and All-India Mean Rainfall (AIMR)- the spatial mean across all grid points. Such differences in variability within the aforementioned homogeneous zones have been studied by (\cite{Azad2010}). However the different aspects of temporal changes and variability, when clustered, cannot be expected to coincide with the clusters formed on the basis of mean rainfall, as observed in \cite{Srinivas2013}. 

Regarding prediction of annual rainfall, an important variable is the sign of year-to-year changes in rainfall. While impacts of rainfall over a season depend on the magnitude and distribution within that season, its change from the previous year is a related variable. Forecasting the change in rainfall from the present year to the next is equivalent to forecasting next year’s rainfall, once the present year’s rainfall is known. The sign of this change is a binary variable, and therefore can be expected to exhibit larger spatial coherence than its magnitude. While this sign alone does not describe the full impacts of rainfall, it represents a compromise between impacts and ability to make forecasts at sub-national scales. Furthermore, the internannual change in AIMR exhibits large mean reversion, and therefore the sign of this change can be predicted with reasonably high confidence. Together, this property of the sign of rainfall change at different spatial scales and their spatial coherence are worth examining. To the best of our knowledge, these properties have not been studied previously. 
Here we find that the sign of year-to-year changes is spatially coherent, but this has different effects from the mean rainfall field. Specifically, clusters describing frequent coincidence of the sign of year-to-year changes differ from the aforementioned clusters defining relatively homogeneous mean rainfall. Therefore they must be examined directly. 

Similarly, it is also important to be able to make forecasts of annual extreme events at local or sub-national scales, i.e. the occurence of years with excess and deficient rainfall. Such years are often associated with floods and droughts respectively, which have very widespread impacts on people's lives and economy in India. We find that there is spatial coherence in the occurrence of local extremes, and clusters can be identified based on such co-occurence. The corresponding clusters tend to differ from the aforementioned clusters formed on the basis of mean rainfall (\cite{Srinivas2013,Gadgil1980}). Identifying grid-level extremes and locations where these coincide with each other is a fundamentally different problem than characterizing variability of large scale patterns using, for example, Empirical Orthogonal Functions as in~\cite{Gadgil1980}. Furthermore, the former problem is not subsumed within that of characterizing spatial patterns of temporal variability, because grid-level extremes need not be correlated with a few large scale spatial patterns of rainfall. Therefore the properties of grid-level extremes and associated clusters must be examined directly.

This paper introduces a systematic approach for identifying homogeneities as well as heterogeneities in year-to-year changes in rainfall as well as annual local extremes. Homogeneities are manifested in spatial coherence, which is an important property of spatiotemporal fields generated by physical processes, and makes possible the identification of relatively homogeneous clusters.  Recently, there has been substantial progress in Data Science and Data Mining, allowing for comprehensive analysis of spatiotemporal datasets (\cite{STsurvey2015}) and extraction of prominent patterns with respect to these homogeneities. We objectively quantify spatial coherence, and use the results to study a number of properties of year-to-year change and annual extremes. The results are applied to identify cases where coherence can be exploited to form significant regionalizations. We analyze 110 years of gridded rain gauge data across India [\cite{Rajeevan2008}], based on concepts of spatiotemporal data mining.  Heterogeneities are manifested in the property that on larger scales there are substantial differences in statistics that also lead to differences from AIMR.

The overall message is threefold. First, spatial heterogeneities are substantial, involving both inter-region differences and differences from the all-India spatial mean. These heterogeneities must be taken into account when considering both year-to-year rainfall changes and extreme rainfall. Second, both these features involve significant spatial contiguities, and hence for both features it is possible to find homogenous spatial clusters. Third, the sign of inter-annual difference is reasonably predictable, and predictability at grid-level improves when combined with national-level prediction. 

\section{Data, Notations and Definitions}

In this work we analyze observations from rain gauges, maintained by the Indian Meteorological Department (IMD), and processed to a $1^{\circ}-1^{\circ}$ grid, comprising 357 spatial indices, for the period 1901-2011 (\cite{Rajeevan2008}). Monthly and annual means are considered. In case of month-wise analysis, each data-point has form $X^{s}_{m}(y)$ where $s$ denotes the spatial location, $m$ the month and $y$ the year. With annual means, each data-point has form $X^{s}(y)$. By averaging across locations (spatial mean), we get All-India Mean Rainfall (AIMR) , denoted by $X(y)$.

\section{Phase of Annual Rainfall}

\subsection{Definition of phase, National-phase, local phase}

Annual rainfall time-series $X^s$ at individual grid-locations indexed by $s$, as well as AIMR, exhibit considerable variability. One property of this variability is mean reversion. In case year $t$ experiences more rainfall than previous year $(t-1)$, then year $(t+1)$ is likely to experience less rain than year $t$. This is partly related to the Tropical Quasi Biennial Oscillation [\cite{Iyengar2005}]. 

The change in rainfall from year to year is an important variable, and directly related to forecasting the next year's rainfall once the present year’s value is known. Furthermore, since the changes occur heterogeneously and do not take the same sign uniformly over India, we identify clusters where these changes frequently coincide. For analyzing this, we define a variable called the location-wise Phase as $P^s(t)=sign(X^s(t)-X^s(t-1))$. For AIMR, corresponding Phase is $P(t)=sign(X(t)-X(t-1))$. For individual locations and AIMR, the time-series of phase $P^s(t)$ and $P(t)$ mostly alternates between $1$ and $-1$ indicating mean reversion. The limitation of phase as defined here is that it does not measure the magnitude of rainfall changes. However its usefulness lies in its higher spatial coherence than changes in rainfall magnitude, as it is a binary variable, and hence amenability to forecasting.

\subsection{Relation between Local and National Phases}
Despite the large scale of monsoon systems, there is coherence in phase over the Indian landmass. This is partly a consequence of the discretization involved in defining phase, where only the sign of rainfall change from year-to-year is considered. Here we identify locations with high probability of having either the same or the opposite phase as that of AIMR, i.e. the national phase. For each location $s$, we compute the set $\{t: P^s(t)=P(t)\}$, including years where local phase agrees with national phase, and denote its cardinality as $PC(s)$. Hence $PC(s)$ describes the number of years when the grid-location agrees in phase with the national phase.

The mean of $PC(s)$ across grid-locations is 70, i.e. locations on average agree in 70 years (out of 110 years of phase data) with the national phase. This is just one effect of the spatial coherence of phase, which therefore becomes easier to predict than the spatial pattern of rainfall. The histogram of relative frequency, $PC(s)$/110, is shown in Figure 1. The figure also identifies locations where $PC(s)$ is unusually high, corresponding to frequent agreement with national phase; and low, corresponding to frequent disagreement. Central and Western India agree with the national phase with high frequency, whereas locations on the South-Eastern coast frequently disagree. Locations with frequent agreement or disagreement (with the national phase) are where the direction of local change can be predicted with high probability based on the spatial-mean forecasts alone.

\subsection{Effect of scale change}
To make analysis more robust to fluctuations at the grid scale, we estimate $Y^s(t)$, the mean of $X^s(t)$ across the 9 grid locations centered at $s$, ignoring locations outside the Indian landmass. The previous analysis is repeated for these so-called “1-hop neighborhoods”. The results are shown in Figure 1. It shows similar results as the previous analysis; however phase in this case is more spatially coherent (Figure 1), with larger contiguous regions frequently having the same phase.

\begin{figure}
	\centering\includegraphics[height=7pc,width=28pc]{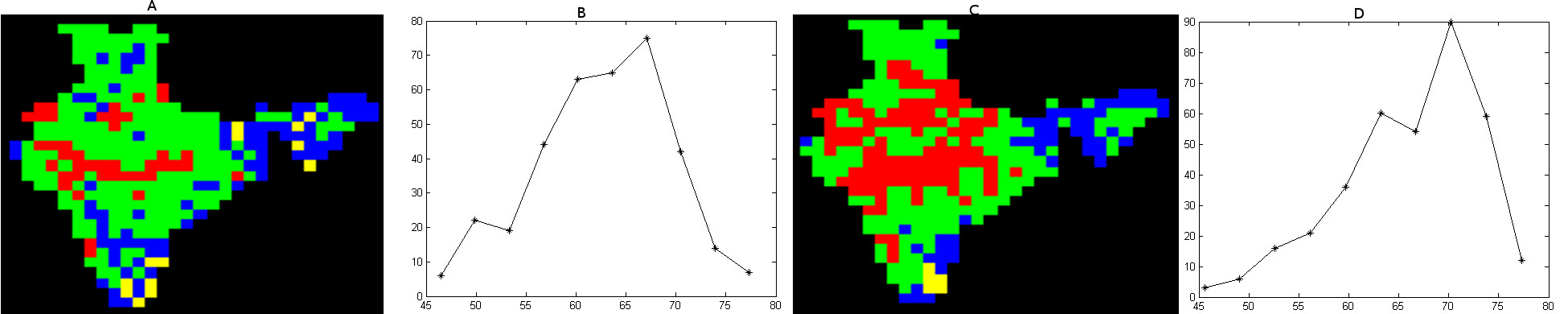}
	\caption{Relation between local and national phase. Most locations have the same phase as the national phase. A: Individual grid-locations grouped according to $PC$/110, their relative frequency of conforming with All-India Mean Phase across the 110 years. (Red: over 70\%, Green: 60-70\% years, Blue: 50-60\%, Yellow: under 50\%. B: the histogram of $PC$/110, in percentages. C: Same as A, but using mean of 1-hop neighbourhoods (defined in Section 3.3) at each location. D: histograms corresponding to C.}
	\label{fig1}
\end{figure}

\subsection{Predictability of Phase}
Indian Meteorological Department (IMD) attempts to predict Indian Summer Monsoon Rainfall (ISMR) each year, based on various meteorological variables. The magnitude of AIMR is difficult to predict, but its phase may be easier to predict, since phase is only a binary quantity and also shows strong mean-reverting behavior due to Quasi-Biennial Oscillation (QBO). In this work, we study its predictability by exploiting only this property, without considering any remote teleconnection effects (such as Sea Surface Temperatures over Pacific and Indian Oceans). Including these will undoubtedly improve the predictability of phase, but that is beyond the scope of this paper. 

Based on the dataset, we make estimates of the conditional distribution $p(P(t)|P(t-1))$. We find that $p(P(t)=1|P(t-1)=-1)=0.64$ and $p(P(t)=-1|P(t-1)=1)=0.68$. This shows that phase of AIMR in any year can be predicted with reasonable confidence by simply considering the phase of AIMR in the previous year. 

Next, we look at the predictability of phase at grid-level. By studying $PC$, we have already studied the probability $p(P^s(t)|P(t))$, which is around 0.62 whenever $P^s(t)=P(t)$. We now evaluate $p(P^s(t)|P^s(t-1))$ for each location $s$, and find that on average (across locations), $p(P^s(t)=1|P^s(t-1)=-1)=0.66$, $p(P^s(t)=-1|P^s(t-1)=1)=0.66$. This means that the mean-reverting property of phase is present at grid-level too, with same strength as in case of AIMR. However, the forecast of AIMR can be improved based on additional variables and remote teleconnection effects, which may not be possible at the grid-level. Therefore, we study how the predictability of grid-level phase can be improved by conditioning on AIMR phase. For this, we study the quantities $p(P^s(t)|P^s(t-1),P(t-1))$ and $p(P^s(t)|P^s(t-1),P(t))$. We find that in about 300 (out of 357 in total) locations, incorporation of national phase of the current year increases this probability i.e. $p(P^s(t)=1|P^s(t-1)=-1,P(t)=1)>p(P^s(t)=1|P^s(t-1)=-1)$ and $p(P^s(t)=-1|P^s(t-1)=1,P(t)=-1)>p(P^s(t)=-1|P^s(t-1)=1)$. Furthermore, in 209 locations, incorporation of national phase of the previous year increases this probability i.e. $p(P^s(t)=1|P^s(t-1)=-1,P(t-1)=-1)>p(P^s(t)=1|P^s(t-1)=-1)$ and $p(P^s(t)=-1|P^s(t-1)=1,P(t-1)=1)>p(P^s(t)=-1|P^s(t-1)=1)$.

Thus we find that, due to the mean-reverting property, both AIMR phase and grid-level phase in any year can be predicted from its previous year's value with reasonable confidence, and the predictability at grid-level increases when combined with national-level information about phase.

\section{Positive and Negative Extremities}
Extreme rainfall can cause floods and droughts, with significant impacts. IMD predicts spatial mean rainfall (AIMR), and included in this methodology is the forecasting of extreme years with respect to seasonal mean rainfall at the national scale. However local extremes can be even more consequential. We consider the spatial association between local extremes, in addition to their association with AIMR, because we would like to explore the extent to which the incidence of local extremes can be inferred (and hence probabilistically forecast) from extremes of AIMR.

\subsection{Spatial, Local, Locational Extremities}

The long-term mean AIMR across years is denoted by $\mu_X$, and corresponding standard deviation by $\sigma_X$. Similarly, at locations $s$, long-term mean rainfall across years differs by location and its mean is $\mu_X(s)$ and standard deviation $\sigma_X(s)$. We examine Positive and Negative Extremities (PEX, NEX) at different spatial scales. 

At the national scale, years of Spatial Positive Extremity are defined as $\{t: X(t) >\mu_X+\sigma_X\}$, and years of Spatial Negative Extremity as $\{t: X(t) <\mu_X-\sigma_X\}$.  For location $s$, Local Positive Extremity years are defined in relation to location-specific statistics, comprising $\{t: X^s(t)> \mu_X(s)+\sigma_X(s)\}$. Similarly, Local Negative Extremity years comprise $\{ t: X^s(t)< \mu_X(s)-\sigma_X(s) \}$.

Next, we define two features called Locational Positive and Negative Extremities. Each year, some locations have local PEXs and others local NEXs. In some years, an unusually high number of locations simultaneously have local PEXs or NEXs. These years are defined as Locational Positive Extremity Years or Locational Negative Extremity Years respectively. These need not coincide with Spatial Extremity Years, as they are defined differently and involve widespread occurrence of local extremes; in contrast with large deviations in the spatial mean. 

We define $NF(t)$ and $ND(t)$ as the number of locations with local PEXs and NEXs respectively in year $t$. Respective means of these variables, across years, are $\mu_{NF}$ and $\mu_{ND}$, and their standard deviations are $\sigma_{NF}$ and $\sigma_{ND}$. Then Locational PEX Years are the set $\{t: NF(t)> \mu_{NF}+ \sigma_{NF} \}$, while Locational NEX Years comprise $\{t: ND(t)> \mu_{ND}+ \sigma_{ND} \}$.

\begin{figure}
	\centering\includegraphics[height=7pc,width=28pc]{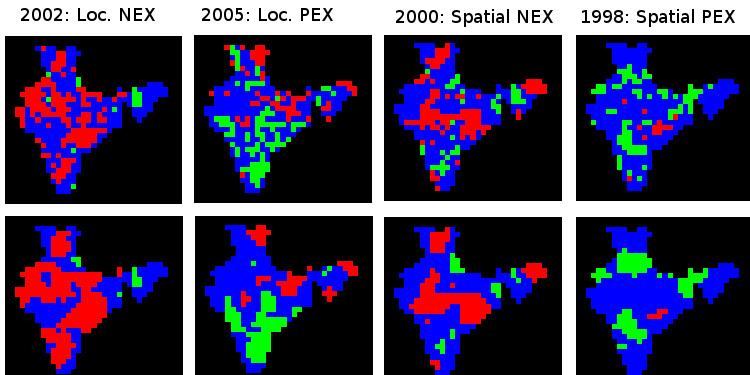}
	\caption{Relation between local and national rainfall extremes in 4 representative years. While local extremes corresponding to the national-scale extreme are more common, opposite extremes also occur, and this is a manifestation of spatial heterogeneity. Locational extremes exhibit more widespread local extremes of the corresponding type than spatial-mean extremes. Colors indicate Local NEXs (red), Local PEXs (green) or normal rainfall (blue). The years are 2002 (Locational NEX), 2005 (Locational PEX), 2000 (Spatial NEX) and 1998 (Spatial PEX) -from left to right. Upper panels indicate grid-specific analysis with IMD data, whereas lower panels repeat the analysis with annual rainfall at each grid-point replaced by the mean of its 1-hop neighbourhood.}
	\label{fig2}
\end{figure}

\subsection{Relation between National and Local Extremes}

As can be expected, locational PEX years turn out to have simultaneously many Local PEXs, because these years are defined as such, i.e. involving an unusually large number of local PEXs. But many locations also experience normal local rain, or even local NEXs, during these years. Analogous behavior is seen during Locational NEX years. This is a manifestation of the heterogeneity of rainfall. The average number of locations having local PEXs is: 113 during locational PEX years, 47 during normal years, and 27 during locational NEX years. Similar statistics are found for mean numbers of local NEXs. Figure 2 illustrates the situation in 4 representative years.

Locations depart in their extreme behavior from the national-scale. Some locations have Local NEXs during several Locational PEX years, or local PEXs during several Locational NEX years. As stated earlier, this is a manifestation of spatial heterogeneity. For this analysis, we estimate the probability of a local PEX/NEX event at each location, conditioned on PEX/NEX events at the national scale. We define random variable $T^s$ as the “year type” at location $s$, which can take values 1 (normal year), 2 (Local PEX), 3 (Local NEX). Similarly we define $T$ as the “year type” for the national scale, which is either 1 (normal year), 2 (Locational PEX), or 3 (Locational NEX). We estimate conditional distributions $p(T^s|T)$ for each $s$. These describe the conditional probability of local year type, given the year type at all-India level. 

The results are illustrated in Figure 3. Only about 60 of the 357 locations have $p(T^s=2|T=2)$ or $p(T^s=3|T=3)$ above 0.4. We also observe that there are some locations where $p(T^s=2|T=3)$ or $p(T^s=3|T=2)$ are significant albeit small. These results suggest that there are only few locations at the grid-scale having substantial probability of conforming in any given year to such national scale extremes, and even there the probabilities are not high. The locations with reasonable correspondance are found to be concentrated along western, central and south-western parts of India, while those with low or negative correspondance are mostly on the eastern side. This indicates that making consistent predictions of grid-scale extremities based on national-scale forecasts alone is not possible, because the national scale extremes do not correspond to frequent repeated incidence of grid-level extremes. What national scale extremes entail at the local scale (i.e. grid-level), if not extremes, is considered next. 
\begin{figure}
	\centering\includegraphics[height=7pc,width=28pc]{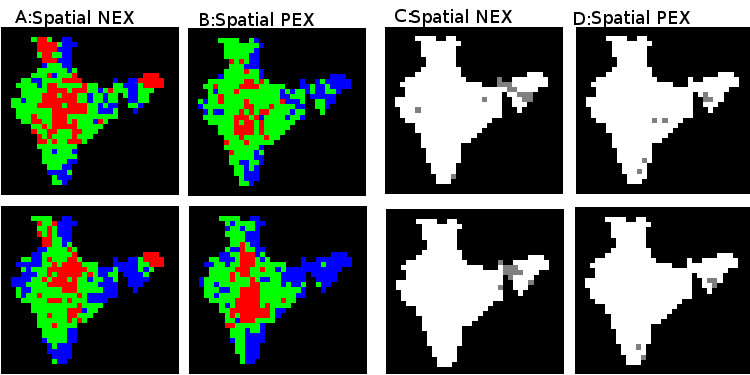}
	\caption{Conditional probabilities of grid-level extremes, given the occurrence of national extremes. These conditional probabilities are not high, and therefore consistent forecasting of grid-level extremes based on national-scale forecasts alone is not possible. A: locations that have Local NEXs in at least 40\% (red) or 20\% (green) of the All-India Spatial NEX years (Blue: Below 20\%). B: locations that have Local PEXs in at least 40\% (red) or 20\% (green) of the All-India Spatial PEX years (Blue: Below 20\%). C: locations that have Local PEXs in at least 20\% (grey) of the All-India Spatial NEX years. D: locations that have Local NEXs in at least 20\% (grey) of the All-India Spatial PEX years. Upper panels indicate grid-specific analysis, lower panels repeat the analysis with annual rainfall at each grid-point replaced by the mean of its 1-hop neighborhood.}
	\label{fig3}
\end{figure}

\section{All-India Extremes and Local Phases}
We now analyze the association between the concepts defined above, namely phase and extremity. We do this because, as discussed previously, the phase is an important variable that exhibits coherence and therefore amenable to prediction at the regional scale. Furthermore, we have already seen through conditional distributions $p(T^s|T)$, illustrated in Figure 3, that local extremities and all-India extremities are not highly correlated. Correlation between local (i.e. grid-level) phase and All-India phase is somewhat higher (Figure 1). We therefore would like to understand precisely what, if anything, national-scale extremities entail at local scales. 
We consider correlations between local phase and all-India extremities, by estimating conditional distributions $p(P^s|T)$. Figure 4 identifies locations where $p(P^s=1|T=2)$ and $p(P^s=-1|T=3)$ are more than 0.7, i.e. where rainfall is likely to increase (compared to the previous year) in the years of all-India PEX, and where it is likely to decrease in the years of all-India NEX. These locations are significant in number (137 and 84 respectively), and distributed all over the main landmass. Thus, there is a significant correlation between all-India extremities and local phase. This property implies that forecasts of strong positive or negative extremities, at the national scale, might be utilized for high probability forecasts of phase in many regions. Hence, although local forecasts of extremities are difficult, corresponding forecast of phase is more feasible; and furthermore, as described below, the probability of correct forecasts increases during years having spatial-mean extremes, when the conditional probabilities of the corresponding phase at the grid-level are higher.

We have already explored the relation between grid-level phase $P^s(t)$ and spatial-mean phase $P(t)$ through the conditional probabilities $p(P^s(t)|P(t))$. We now examine the quantity $p(P^s(t)|P(t),T(t))$, i.e. whether the relation between national and local phase is stronger during years with extremes in the spatial-mean rainfall. It turns out that for all 357 locations, $p(P^s(t)=1|P(t)=1,T(t)=2)\ge p(P^s(t)=1|P(t)=1)$ and $p(P^s(t)=-1|P(t)=-1,T(t)=3)\ge p(P^s(t)=-1|P(t)=-1)$. This means that in spatial-mean extreme years, all locations are more likely to follow the spatial-mean phase than in normal years. Averaged across all locations, the probability of conforming to the spatial-mean phase is about 0.66 in both spatial-mean PEX and NEX years, and about 0.62 otherwise.

\begin{figure}
	\centering\includegraphics[height=7pc,width=28pc]{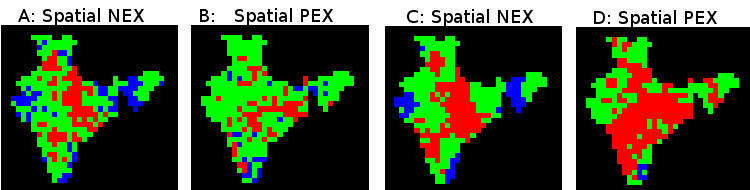}
	\caption{Conditional probabilities of local phase, given the occurrence of national extremes. High probability forecast of local phase is possible, conditioned on the occurrence of national extremes. The probabilities are higher than would be possible with knowledge of spatial-mean phase alone. A: locations in negative phase (decrease in annual rainfall) in at least 70\%(red) or 50\%(green)  of Spatial NEX years (Blue: Below 50\%). B: Locations in positive phase (increase in annual rainfall) in at least 70\%(red) or 50\%(green) of Spatial PEX years (Blue: Below 50\%). C: Same as A with annual rainfall at each grid-point replaced by mean of its 1-hop neighborhood. D: Same as B with annual rainfall at each grid-point replaced by mean of its 1-hop neighborhood.}
	\label{fig4}
\end{figure}

\section{Spatial Coherence}

Geophysical phenomena are spatially coherent, with generally higher correlations at smaller distance. Spatial coherence is not merely another property of the dataset but also helps point the way towards the possibility of forecasting as well as methods for doing so. Specifically where a variable is more coherent, it indicates the possibility of forecasting using less information, and forecasts can be made for several locations simultaneously. Therefore we consider spatial coherence of local extremities and local phase by answering the following questions:
\begin{enumerate}
	\item Are adjacent locations likely to be in the same phase every year?
	\item Is the agreement or disagreement of the phase at grid-level with the spatial-mean phase spatially coherent? 
	\item If a location has a local extremity (PEX/NEX) in any year, are its neighbors more likely to have the same extremity in that year?
	\item Does coherence of local extremities increase during national-level extremities?
\end{enumerate}
Understanding such properties helps us develop a conceptual basis for clustering of phase and extremes, which is described in the following section. To answer these quantitatively, we define two measures of spatial coherence with respect to any property $P$:
\begin{enumerate}
	\item For locations having property $P$, the Mean Number of 1-hop Neighbors (MNN), with 1-hop neighborhood defined in Section 3.3, also having $P$
	\item Mean Connected Component Size (MCCS) of a graph where each vertex represents a grid location, and two vertices are joined by an edge if and only if they are 1-hop neighbors on the grid and both have property $P$. 
\end{enumerate}

These two measures elicit different but complementary aspects of coherence. The MNN describes only the mean properties of the isotropic 1-hop neighbourhoods, whereas the MCCS considers coherence over more general anisotropic domains. The conclusions from applying these two measures could differ, but we use them in conjunction to evaluate the following properties: Being in Positive Phase in a year (PP), Being in Negative Phase in a year (NP), Agreement with All-India Phase in a year (AP), Disagreement with All-India Phase in a year (DP), Having a local NEX in a year (LN), Having a local PEX in a year (LP), Having a local NEX in a year of spatial/locational PEX/NEX (LN-SP, LN-SN, LN-LN, LN-LP) and having a local PEX in a year of spatial/locational PEX/NEX (LP-SP, LP-SN, LP-LN, LP-LP). These properties are evaluated for each location, and both the measures of spatial coherence (MNN and MCCS) introduced above are calculated for each property. Results are shown in Table 1.

\begin{table}\label{tab:tab1}
	\tiny
	\caption{Spatial Coherence for the different properties introduced in Section 5}
	\begin{tabular}{| c | c | c | c | c | c | c | c | c | c | c | c | c | c | c |}
		\hline
		Measure	& PP & NP & AP & DP & LN & LP & LN-SP & LN-SN & LN-LP & LN-LN & LP-SP & LP-SN & LP-LP & LP-LN\\
		\hline
		MNN & 4.98 & 4.87 & 6.69 & 5.60 & 3.86 & 3.90 & 2.73 & 5.2 & 2.83 & 5.44 & 4.91 & 3.40 & 5.18 & 3.49\\
		MCCS& 2.17 & 2.14 & 2.75 & 1.56 & 1.15 & 1.15 & 1.03 & 1.39 & 1.05 & 1.46 & 1.37 & 1.06 & 1.43 & 1.06\\
		\hline
	\end{tabular}
\end{table}

The main messages emerging from results in Table 1 are:
\begin{enumerate}
	\item Local Phase is spatially coherent, because mean number of neighbors for properties PP and NP (being in positive and negative phase respectively) are about 5 (which is about 60\% of the neighborhood size of 8). Therefore if a location is in positive or negative phase, about 60\% of its neighbors have the same phase. However on averaging across grid locations, the mean probabilities of being in positive or negative phase are each close to 50\%. Therefore these results indicate that being in a certain phase is more likely when its neighbors are in the same phase.
	\item Agreement with All-India phase is more spatially coherent than disagreement, because both MNN and MCCS are larger for property AP (agreement with national phase) than for DP (disagreement with the national phase). Therefore if a location agrees with the national phase, the probabilities that its neighbors also agree with the national phase are larger than otherwise.
	\item Local extremities are spatially coherent because the MNN for LP (local PEX in a year) and LN (local NEX in a year) correspond to conditional probabilities of PEX and NEX that are significantly larger than the unconditional probabilities of locations having PEX and NEX respectively. In an average year, only about 14\% locations of India have local PEX (or local NEX), but 3.9 out of 8 neighbors (nearly 50\%) of a local PEX (or NEX) also have local PEX (or NEX).
	\item Local Extremities of any sign are more spatially coherent in years of All-India extremities of same sign; and less spatially coherent in years of All-India extremities of opposite sign. This is inferred from MNN and MCCS being smaller for property LN-SP (local NEX during spatial PEX) than for LN-SN (local NEX during spatial NEX), and likewise for property LP-SN (local PEX during spatial NEX) than for LP-SP (local PEX during spatial PEX). Similar relations hold in case of locational extremities. For example in a spatial mean NEX year, there is larger coherence around locations with local NEXs as manifested by larger MNN and MCCS. 
\end{enumerate}

In summary, for phase, the local phase is coherent and agreement with AIMR phase is more coherent. This raises the possibility of forecasting phase through clustering, despite the substantial heterogeneities in rainfall. Furthermore, despite the generally weaker association between local and national extremes discussed previously, local extremes exhibit coherence and this raises the possibility of using further information to forecast coherent clusters where extreme events can be used. Because this coherence is higher during spatial mean extremes of the same sign, such cluster-based forecasting would be more plausible for sub-national scale extremes having the same sign as the national-level extreme.

\section{Discovering Homogeneous Regions by Clustering}
Motivated by this discovery of spatial coherence, we now try to identify small but spatially contiguous sets of grid-locations with homogeneous behaviour with respect to phase and local extremes. Earlier, different criteria such as mean and variability of rainfall have been used to identify such clusters (\cite{Srinivas2013,Gadgil1980}). Here we extend such analyses to include phase and extremities as defined above. 

\subsection{Similarity Measures and Spectral Clustering}
Clustering techniques, well-known in Data Mining, seek to partition data according to predefined measures of similarity. Each partition is called a \emph{cluster}. We use Spectral Clustering (\cite{Ng2002}) to partition the grid into relatively homogeneous clusters. This method takes as input an $N\times N$ matrix $S$, where $N$ is the number of data-points and each entry $S(a,b)$ encodes a measure of similarity between the datapoints indexed by $a$ and $b$. This similarity measure is application-specific. We are also required to specify the number of clusters $K$. Clustering algorithms assign to each data-point a cluster index in $\{1,\dots,K\}$, with the points having the same index comprising a cluster. Generally points assigned to the same cluster are expected to be more "similar" to each other than to other points, with respect to the measure of similarity in $S$.

The goal of clustering with respect to local phase and extremes is to identify regions at the sub-national scale where these properties are similar. Hence we smooth the data by estimating means over 1-hop neighborhoods at each grid-location, before applying the clustering algorithms below.

\subsection{Identification of Co-occurring Phases}
To generate clustering according to phase, for every pair of locations $(a,b)$ we compute $S(a,b)= |\{t: P^a(t)=P^b(t)\}|$, i.e. $S(a,b)$ is the number of years when $a$ and $b$ have same phase. Thus the measure of similarity is related to the tendency of the locations to be in the same phase. By specifying a small number of clusters, we identify coherent regions that coincide in phase much of the time. Although the spectral clustering algorithm is not biased to manifest spatial contiguity of clusters, spatial neighbors often appear in the same cluster. Not all clusters formed by the algorithm have high intra-cluster similarity with respect to $S$ defined above. In Figure 5, we only show the clusters having high intra-cluster similarity, such that locations within any one of the selected clusters coincide in phase in a large number of years. These are the clusters for which comparatively higher probability forecasts of phase might be possible.

\begin{figure}
	\centering
	\includegraphics[height=5pc,width=14pc]{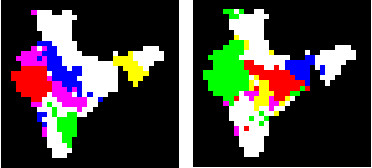}
	\caption{Spectral clustering of locations according to phase, with colors indicating different clusters. Left: Locations in each cluster are in same phase in at least 70\% of years. Right: Locations in each cluster are in same phase in at least 70\% of the all-India PEX and NEX years.}
	\label{fig5}
\end{figure}

\subsection{Identification of Co-occurring Local Extrema}
To identify locations simultaneously experiencing either local positive or negative extremes, we define similarity measure for each pair $(a,b)$ as $S_{F}(a,b)= |\{t: T^a(t)=2, T^b(t)=2\}|$ for PEX (the number of years when both $a$ and $b$ have local PEX), and $S_{D}(a,b)= |\{t: T^a(t)=3, T^b(t)=3\}|$ for NEX (the number of years when both $a$ and $b$ have local NEX). Using these as the similarity matrices for spectral clustering with 10 clusters, we find in case of strong intra-cluster similarity the clusters of locations frequently experiencing local PEXs (or NEXs) in the same years. Locations in any one of the clusters in Fig 6A have simultaneous local PEX in at least 8 (about $50\%$) of their local PEX years. In Fig 6B we show clusters of locations which have simultaneous local PEX in at least 3 (about 20\%) all-India PEX years.
For this analysis, we use similarity matrix $S_{SF}(a,b)= |\{t: T^a(t)=2, T^b(t)=2, T(t)=2\}|$.  Locations in any of the clusters in Fig 6C have simultaneous local NEX in at least 6 (about $40)\%$ of their local NEX years, and locations in the clusters of Fig 6D have simultaneous local NEX in at least 3 (about 20\%) all-India NEX years. For Fig 6D, the similarity matrix used is $S_{SD}(a,b)= |\{t: T^a(t)=3, T^b(t)=3, T(t)=3\}|$.

Such large and spatially contiguous clusters could not have arisen if the occurrence of local NEX or PEX were independent across the locations within the clusters, and are one manifestation of the spatial coherence described previously. Although we are smoothing the results using means over 1-hop neighborhoods before applying the spectral clustering algorithm, we note that similar (but noisier) clusters are obtained if the algorithm were implemented on the data without smoothing. 

However, the threshold probability for including these clusters in the figure is much lower than that of phase. With a higher threshold of $60\%$, very few clusters survive. Therefore while there is coherent behavior in extremes across many locations, the conditional probability of NEXs across these clusters given a spatial mean extreme is small. Therefore high probability forecasts of local extremes, or extremes across individual clusters, cannot rely on the spatial mean forecast alone. Even the clusters that are not conditionally dependent on spatial mean extremes (in Fig 6A and 6C) do not have simultaneous local extremes with as high probability as in the case of the clusters involving phase in Fig 5. This indicates that forecasting extremes at the sub-national scale is fundamentally more difficult than forecasting phase.  

\begin{figure}
	\centering
	\includegraphics[height=6pc,width=28pc]{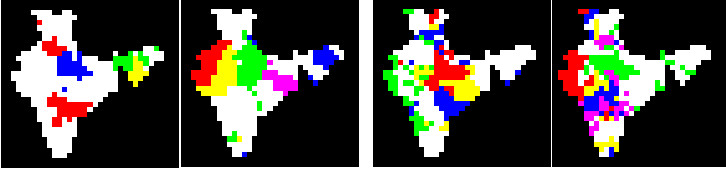}
	\caption{Spectral clustering of locations according to co-occurrence of local extremes, with colors indicating different clusters. 6A: Locations in each cluster have local PEXs simultaneously in at least 8 years (50\% of their local PEX years). 6B: Locations in each cluster have local PEXs simultaneously in at least 3 years of all-India spatial PEX (20\% of spatial PEX years). 6C and 6D: the same analysis, repeated for local and all-India spatial NEX}
	\label{fig6}
\end{figure}

\section{Conclusions}

Forecasting rainfall received in any year is important for India's economy, especially for millions of workers directly dependending on rainfed agriculture for their livelihoods. Currently, IMD makes forecasts of seasonal rainfall at the national scale, and these seasonal-mean forecasts do not carry direct implications for individual regions. However impacts are mainly felt through events occurring at local and sub-national scales. 

Making regional or grid-level forecasts is far more difficult due to the heterogeneity of rainfall, but our findings show that some weaker forecasts (related to phase) might be possible at smaller scales, and that could furthermore be largely based on AIMR forecasts. 

This is mainly because the phase, being a binary quantity, exhibits larger spatial coherence and larger association with the all-India phase, and therefore is more amenable to prediction based on the AIMR change. We showed here that, despite the heterogeneity, local phase has a high probability of following the national phase; and this probability is higher during extreme years. Furthermore, the identification of coherent clusters in which the phase coincides with high frequency raises the possibility of the forecasting of phase within these clusters. 

The treatment of phase has more general consequences for understanding variability; for example one could think of clusters with mostly the same phase as being those that vary together (after neglecting the magnitude of changes). Understanding the phase and its behaviour is important because the knowledge of phase indicates whether rainfall in the following year will be more or less than the present year, and this type of comparison might be important for adaptation to rainfall variability. 

The results presented here also showed that, corresponding to the mean-reversion of AIMR, its phase is reasonably predictable. Therefore, years with negative phase have high probability of being followed by positive phase years; and vice versa. This also carries over to the grid-level, despite the spatial heterogeneity in phase. Together, the mean reverting character of large-scale rainfall and the discretization involved in defining the phase make this variable more predictable than many of its alternatives. 

Another aspect of the paper is in understanding the properties of years experiencing extreme rainfall. Extreme rainfall is important for impacts, and local and sub-national extremes play the largest roles. Statistical forecasting of the monsoon does not extend to seasonal forecasts at the grid level, but it is important to understand the relation between local and national extremes, and whether local extremes can be forecast. Here we consider grid-level extremes, and their relation with spatial-mean extremes. The results show that grid-level as well as regional extremes do not follow the national extremes with high probability. This implies that national-level forecasts alone cannot be deployed to infer or predict grid-level extremes with high confidence. 

However it was objectively shown that grid-level extremes exhibit spatial coherence. Moreover the coherence increased in the presence of a national-scale extreme of the same sign. This, together with the demonstration that extremes tend to occur in spatially contiguous clusters, raises the possibility of forecasting extremes at the level of individual clusters where extremes tend to coincide. Such forecasts would only be probabilistic, and making such forecasts would require exploiting additional information than merely the occurrence of national-level extremes, such as sea surface temperature patterns.  Beyond such speculation based on the results presented here, the exploration and elaboration of such methodology is outside the scope of this paper.  However it was also shown that such efforts are likely to face intrinsic difficulties because grid-level extremes within relatively homogeneous clusters do not coincide with high probability, as shown in the previous section. 

A number of possible regionalizations can be made, some more useful than others. However a necessary condition for the existence of contiguous clusters based on some variable is that the variable should exhibit spatial coherence. By introducing objective measures of coherence that cover both isotropic and anistropic cases, we were able to identify some variables for which significant regionalizations are possible. 

Long-term simulations of Indian rainfall are needed to formulate socio-economic and developmental policies, especially in the presence of climate change, and such simulations should be able to capture regional variations accurately. It has been noted that under the influence of climate change, such spatial differences are actually increasing (\cite{Ghosh2009,Ghosh2012}). We have identified additional salient characteristics of spatiotemporal heterogeneities relevant to evaluating simulations by climate models (\cite{Raj2015}). These findings by previous authors of the non-stationarities of extreme rainfall statistics suggest that our present analysis assuming a stationary climate must be extended to consider how the clusters and associated relationships examined here are evolving in time due to climate change.

\section{Acknowledgments}
This research was supported by Divecha Centre for Climate Change, Indian Institute of Science. We are thankful to Dr. J. Srinivasan, Dr. V.Venugopal and Dr. K. Rajendran for their valueable inputs.


\end{document}